%% file: hagiwara.tex
\documentclass[12pt,preprint]{aastex}

\newcommand{\kms}{km s$^{-1}\;$}
\newcommand{\kmss}{km s$^{-1}$}
\newcommand{\km}{km s$^{-1}\;$}

\newcommand{\vlsr}{$V_{\rm LSR}$}

\newcommand{\lsun}{\mbox{$L_{\sun}$}}

\newcommand{\ho}{H$_{2}$O$\;$}

\newcommand{\mb}{mJy beam$^{-1}$}

\newcommand{\hi}{H$_{\rm I}\;$}

\shorttitle{Water maser in NGC 6240}
\shortauthors{Y.Hagiwara}

\begin{document}

%
%
\title{Two Epochs of Very Large Array Observations of Water Maser Emission in the active galaxy NGC 6240}


\author{Yoshiaki Hagiwara\altaffilmark{1,2}}
\affil{$^1$National Astronomical Observatory of Japan, 2-21-1 Osawa, Mitaka, Tokyo 181-8588, Japan}
\affil{$^2$Department of Astronomical Science, The Graduate University for Advanced Studies (Sokendai), 2-21-1, Osawa Mitaka, 181-8588 Tokyo, Japan}



%
\begin{abstract}
  Studies of 22GHz H$_2$O maser emission from the merging galaxy
  \object{NGC 6240} with double nuclei are presented. Two epochs
  of Very Large Array (VLA) observations in the A-configuration in
  spectral-line mode were carried out at 0.1 arcsec resolution by
  covering the redshifted velocity range of $\sim$ 300 \kms with
  respect to the systemic velocity of the galaxy.  The purpose of
  these new observations is twofold: to detect an \ho maser that an
  earlier VLA observation pinpointed in the southern nucleus in 
  the northern nucleus as well to clarify the kinematics of the double
  nuclei, and to understand the origin of the maser in the galaxy.  In
  the second epoch, one velocity feature peaking at \vlsr = 7491.1
  \kmss, redshifted by $\sim$200 \kms relative to the systemic
  velocity, was detected only toward the southern nucleus.  The
  detection of an \ho maser feature at or near this velocity had never
  been reported in earlier observations.  However, including the known
  velocity features at redshifted velocities, no other velocity
  features were observed toward either nuclei throughout these
  epochs.  The maser remains unresolved at an angular resolution of
  $\sim$ $0\arcsec$.1, corresponding to a linear size of less than
  about 45 pc.  The two epochs of VLA observations show that the
  maser intensity is variable on timescales of at least three months,
  while the correlation between the maser intensity and the radio
  continuum intensity is not certain from our data. It is plausible
  that the maser in \object{NGC 6240} is associated with the activity
  of an active galactic nucleus in the southern nucleus.
  Alternatively, the maser can be explained by starforming activity at
  the site of massive starformation in the galaxy.
\end{abstract}
%
%
\section{Introduction}
%
%

\object{NGC 6240} is a luminous infrared merging galaxy with large
far-infrared (FIR) luminosity of $L_{\rm FIR}$ = 3.5 $\times$
10$^{11}$ \lsun\,\citep{yun02}.  Due to this, \object{NGC 6240} has been categorized as a luminous
infrared galaxy (LIRG) with $L_{\rm FIR}$ = 10$^{11-12}$ \lsun, or
almost as an ultraluminous infrared galaxy (ULIRG) with $L_{\rm FIR}$
$>$ 10$^{12}$ \lsun \citep[e.g.,][]{sand88}.
The power source of the large infrared luminosity of U/LIRGs is
generally believed to be starburst activity induced by the
galaxy$-$galaxy interaction or merger and UV radiation from massive
star formation is thought to be reprocessed to FIR radiation
in the dusty environment \citep[e.g.,][]{sand88, ski97}.
Infrared spectroscopy of hydrogen emission lines shows that the origin
of the large FIR luminosity of \object{NGC 6240} cannot 
account for only the starburst activity but can be more naturally
understood in terms of a buried active galactic nucleus (AGN) heating
surrounding dusty components. \citep[e.g.,][]{dep86}.
$Infrared\,Space\,Observatory$  ($ISO$) observations revealed that
the most dominant power source of  ULIRGs including \object{NGC 6240} 
 within the large apertures of $ISO$ is a starburst.
\citep{gen98}.   
$Spitzer\,Space\,Telescope$ observations of \object{NGC 6240} detected a
buried AGN for the first time in the mid-infrared (MIR) band, which suggests
the presence of warm dust ($\sim$ 700 K) surrounding the AGN
\citep{arm06}.  Modeling of the total spectral energy distribution from
near-infrared (NIR) to FIR wavelengths suggests that the fraction of AGN
contribution to the bolometric luminosity in NGC 6240 is approximately
20$\%$$-$24$\%$ \citep{arm06}, which is consistent with $ISO$ studies.
The galaxy hosts two nuclei that are distinctly separated by 1$\arcsec$.5-1$\arcsec$.8,
which depends on observing wavelengths \citep{fri83, col94, bes01, max05}.
The hard X-ray detection of the two point sources, positions of which
coincide with those of the two optical and infrared nuclei, provides
observational evidence for a merging supermassive black hole (SMBH)
scenario \citep{kom03}.  Both nuclei show characteristics of AGNs
based on the detection of neutral Fe K$\alpha$ lines at 6.4 keV due to
reflection from optically thick material \citep{bol03, kom03}, and a
most prominent AGN is situated at the southern nucleus, where the
obscuration is higher.

\object{NGC 6240} is known to show prominent \hi and OH absorption
against the nuclear radio continuum \citep{baa85}.  MERLIN
observations distinguished the different velocity structures at each
of the double nuclei in terms of \hi absorption \citep{bes01}. 
Earlier Very Large Array (VLA) observations in the 1.4 and 1.6 GHz bands \citep{hagi07b,baa07} 
studied both the extended
and compact \hi and OH absorption: the \hi is distributed across the
extended radio continuum structure with a significant concentration
toward the two nuclei, while OH is confined mostly between the
nuclei, closer to the peaks of thermal molecular gas such as CO
\citep[e.g.,][]{naka05}.  Moreover, Very Long Baseline Interferometer
(VLBI) observations with the Very Large Baseline Array (VLBA) \citep{gall04} detected two compact
radio sources near the positions of the two radio nuclei
\cite[e.g.,][]{col94}.  Brightness temperatures of these sources exceed 7
$\times$ 10$^6$ K for the northern component and 1.8 $\times$ 10$^7$ K for the southern component, suggesting that both are
associated with AGNs.

Observations of thermal molecular lines with millimeter
interferometers identified significant concentrations of dense
molecular gas between the double radio/optical/X-ray nuclei
\citep{bry99, tac99,naka05,dai07}, where a significant amount of
H$_2$ emission is detected \citep{ohy00}.  \citet{dai07} updated the
values of dense and warm gas mass in the central 1 kpc of the galaxy
and reported the detection of a dust continuum at submillimeter
wavelengths.

The tentative detection of \ho maser emission in the galaxy was first
reported in 1984 by Henkel et al.. However, the emission was only confirmed
 in the spring of 2001 by two independent measurements
 \citep{hagi02, naka02}.  In early summer 2001, detections
of several velocity features in \object{NGC 6240} were reported
from the measurements at Effelsberg, Greenbank, and Nobeyama
\citep{hagi02,bra03,naka08}.  Follow-up interferometric observations
of these maser features were performed and the location of one of the
maser features was pinpointed in the center of the southern nucleus of
NGC 6240 \citep{hagi03}.

New interferometric observations of the \ho maser intended to observe the
binary AGN in NGC 6240 and aiming at the first detection of the \ho maser
from the northern nucleus are presented here. The results
of these new observations will be useful as a basis for
future detailed studies of the black hole masses and kinematics in
this rare binary SMBH system. Cosmological parameters of $H_{0}$ = 73
\kms Mpc$^{-1}$, $\Omega$$_{\Lambda}$ = 0.73, and $\Omega$$_{M}$ =
0.27 are adopted throughout this article. The luminosity distance to
NGC 6240 is therefore 103 Mpc ($z$=0.0245), and 1 arcsec corresponds to
476 pc in the linear scale.
\section{OBSERVATIONS AND DATA REDUCTION}
Observations of 22 GHz \ho maser emission (6$_{16}-5_{23}$
transition) toward \object{NGC 6240} were conducted using the
NRAO\footnote{The National Radio Astronomy Observatory is
 a facility of the National Science Foundation operated under cooperative
  agreement by Associated Universities, Inc.} VLA in A$-$configuration on 2008 October 3 and 2009 January 12. 
Each observation lasted approximately four hours.  During this period, due to a source of unknown noises associated with using the 12.5 MHz bandwidth
with the VLA correlator, we chose to use the VLA 2AB correlator mode 
 with two 6.25 MHz intermediate frequencies (IFs) at two different
frequency settings. The observations were thus made employing the two
IFs of 6.25 MHz bandwidth with a single polarization, divided into 64
spectral channels, yielding velocity resolutions of 98 kHz (1.32
\kmss).  For this reason, the velocity resolution of our new
observations is narrower than that of earlier
VLA observations in \citet{hagi03} by a factor of two.  Phase-referencing observations were
performed by switching to a nearby phase-referencing source 1658+074,
the position of which is 5$\degr$.4 away from NGC 6240.
Each of the two 6.25 MHz IFs was centered at one frequency setting
and changed to the other frequency setting by shifting the center frequency
during observations. By doing this, we obtained four IF center frequencies,
 centered at \vlsr = 7410, 7480, 7550, and 7620 \kmss, 
which resulted in $\sim$ 300 \kms effective velocity coverage. 
We conducted phase-referencing observations at
one frequency set for three minutes and changed the observation to the other frequency set for three minutes. The observations were thus made cycling
between the two frequency settings so that the maximum velocity coverage
was obtained within the limited bandwidths.
However, due to this time-share between two
frequency settings, sensitivities are a factor of 1.4 worse than what
would be achievable using the two 12.5 MHz IFs.

Amplitude and bandpass calibration was performed using observations of
3C\,286.  The flux-density scaling error relative to assumed standards
is reported by the VLA Observational Status Summary to be less than 1\% in the 22 GHz band \citep{ulv09}.  However, flux scaling
errors of 5\%  are conservatively adopted.  Flux-density
uncertainties thus include noise, the assumed 5 \% uncertainty in
the flux density scale, and other calibration errors such as fitting
errors, added in quadrature.  For the calibrator source, 1658+074, we
obtained a flux density of 1.98 $\pm$ 0.02 Jy.

The data were calibrated and imaged using the NRAO
Astronomical Image Processing Software (AIPS).  After the phase and
amplitude calibrations, the 22 GHz continuum emission of NGC 6240 was
subtracted from the spectral line visibilities using line-free
channels prior to the imaging and CLEAN deconvolution of the maser
emission. The maser in \object{NGC 6240} was thus separated out from
the 22 GHz continuum emission, while the continuum image was made from
single IF channel ("channel 0")  visibilities.

The synthesized beam sizes produced from uniformly and
naturally weighted spectral-line maps were 0.104 $\times$ 0.092
arcsec$^2$ (P.A.=10$\degr$.4) and 0.135 $\times$ 0.082 arcsec$^2$
(P.A.=8$\degr$.6) respectively.  The beam size of a uniformly-weighted
continuum map is 0.097 $\times$ 0.085 arcsec$^2$ (P.A.=10$\degr$.4).  
A naturally-weighted continuum map is shown in Figure~\ref{figure1}.
The rms noises of spectral line maps in uniform and natural weight
were $\sim$ 1.1 \mb and $\sim$ 0.8 \mb per spectral channel
respectively. The rms noises of the continuum maps were $\sim$ 0.3 \mb
(uniform weight) and $\sim$ 0.25 \mb (natural weight).  The optical
velocity definition is adopted throughout this article and the
velocities are calculated with respect to the local standard of rest
(LSR).
%
%
\section{DATA ANALYSIS AND RESULTS}
\begin{figure}
\epsscale{0.9} \plotone{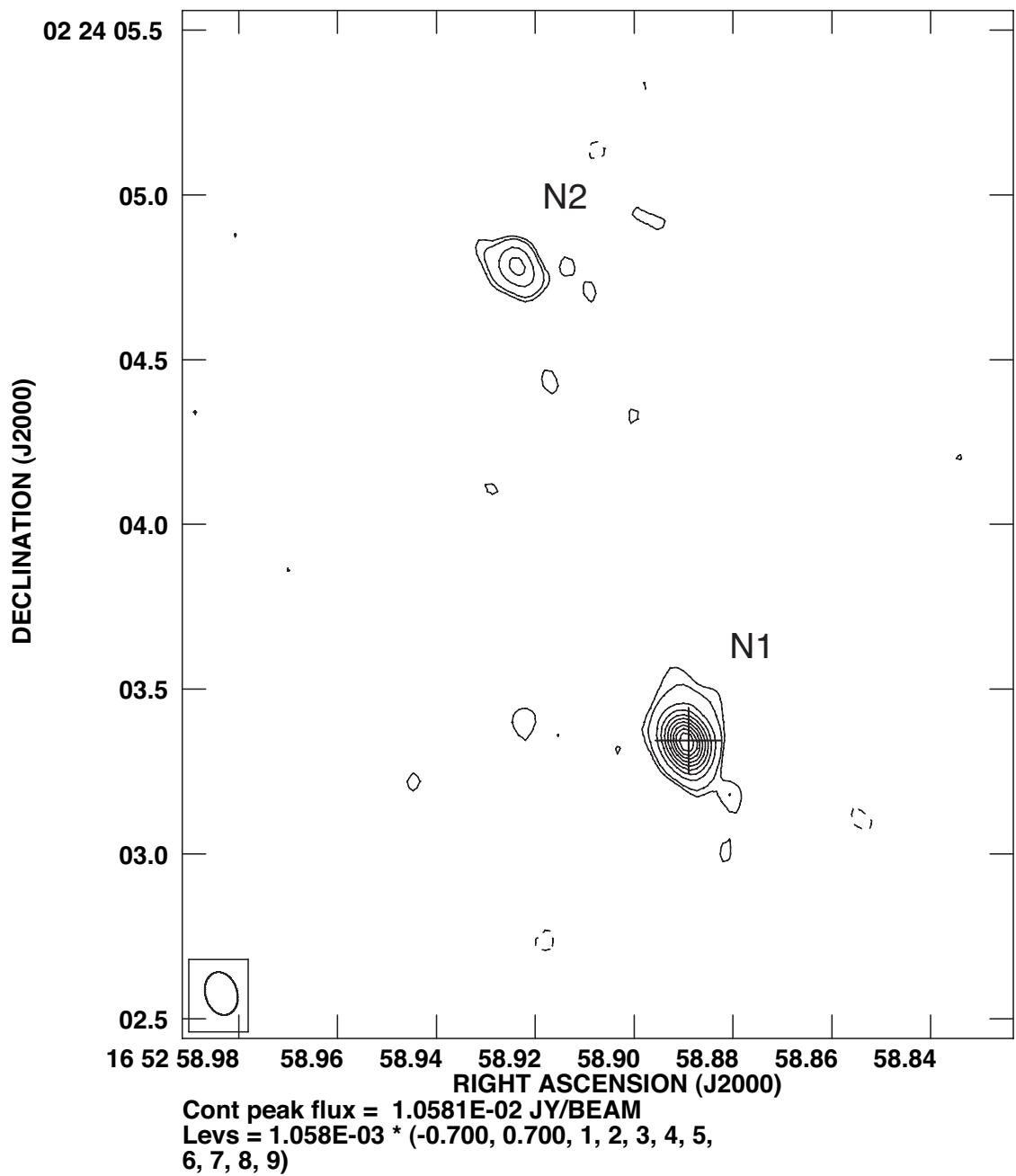}
\caption{22 GHz radio continuum map of NGC 6240, obtained from VLA
  (6.25 MHz bandwidth) in the A-configuration. The two radio nuclei, N1 and
  N2 are labeled.  This map, averaged over two observing epochs on 2008  October 3 and 2009 January 12, is made using natural weighting.
  Contours are -7\%,7\%,10\%,20\%,30\%,40v,50\%,60\%,70\%,80\%, and 90\% of the peak flux
  density of 10.6 \mb, the synthesized beam size of 0$\arcsec$.132 $\times$ 0$\arcsec$.096, with 
  a position angle of 16$\degr$.5.
 The synthesized beam (FWHM) is plotted in the
  left bottom corner of this map. The position of the detected \ho
  maser peaking at \vlsr=7491.1 \kms is denoted by a
  cross. \label{figure1}}
\end{figure}

The data obtained from the two epochs of VLA observations are summarized
in Table~\ref{table1}.  A naturally weighted 22 GHz continuum map
produced from the combined data of the two epochs is presented in
Figure~\ref{figure1}. The 22 GHz continuum map consists of the two
radio sources, both coinciding with the radio nuclei
obtained at lower frequencies with similar angular resolution
\citep[e.g.,][]{car90, col94, bes01}. However, extended
components \citep[e.g.,][]{col94, baa07} seen at lower VLA resolution maps in
the literature were resolved in our continuum map due to the lack of short
spacings of the A-configuration.  Hereafter, we follow the naming convention of
radio components in \object{NGC 6240} by \citet{col94} for the nuclei N1 (the southern nucleus) and N2 (the northern nucleus).

\begin{figure}
\epsscale{1} \plottwo{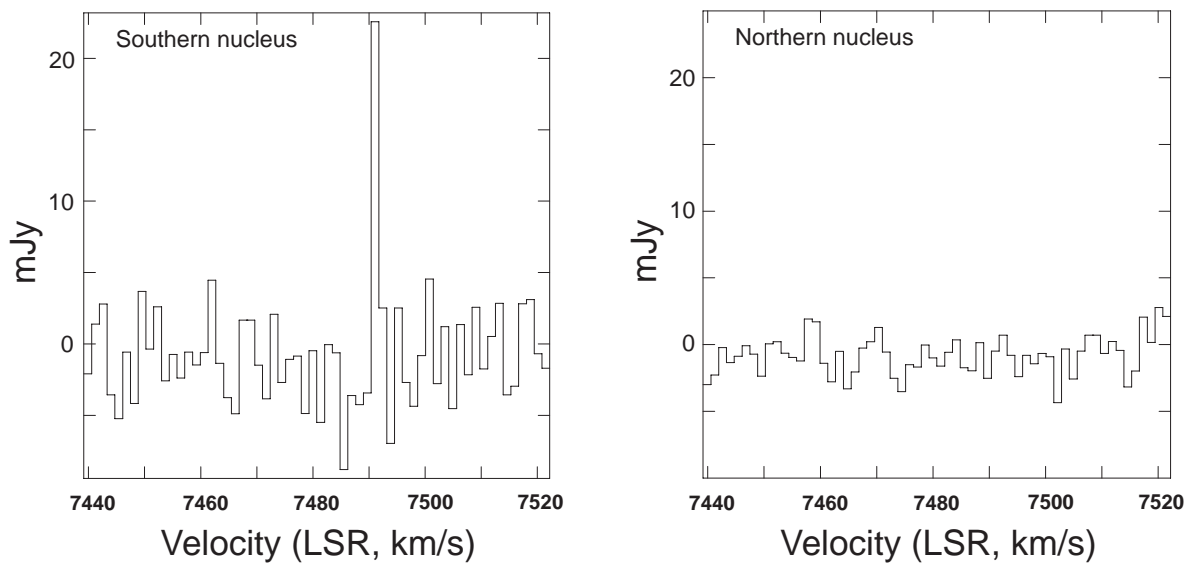}{f2b.eps}
\caption{Left: \ho maser spectrum (\vlsr $\simeq$ 7440--7520 \kmss)
  for the southern nucleus N1, observed with VLA-A on 2009 January 12.
  The spectrum was produced from naturally weighted spectral-line
  velocity maps. The detection of the maser was at \vlsr = 7491.1
  \kmss. The isotropic luminosity calculated from this feature is
  $\sim $ 7 \lsun. Right: VLA spectrum with natural weighting for
  the northern nucleus N2, taken in the same velocity span as the
  maser spectrum. This spectrum is made by averaging data over the two
  observing epochs. No \ho maser was detected toward the northern
  nucleus throughout the two epochs.  \label{figure2}}
\end{figure}

\ho masers in the galaxy were sought for each nucleus in the LSR
velocity range of 7370$-$7660 \kmss, where \ho maser features were
detected in earlier single-dish observations (see Table~\ref{table2}).
In the first epoch of 2008 October 3, no \ho maser was detected with
a 5$\sigma$ noise level of $\sim$ 4 mJy. The maser was detected at
\vlsr= 7491.1 $\pm$ 0.2 \kms on 2009 January 12 in the second epoch,
although the maser remains unresolved at an angular resolution of
$\sim$ 0$\arcsec$.1 or 47.6 pc. There is no doubt that the maser
varied within about three months in the period between October 2008
and January 2009.  The position of the only maser detected in this
program is $\alpha$(J2000): 16$^{\rm h}$52$^{\rm m}$58$\fs$889,
$\delta$(J2000): +02$\degr24\arcmin 03\arcsec$.34. (Position
errors without consideration of systematic errors are $\approx$
0$\arcsec$.007.) Throughout the two observing epochs, in the
spectra averaged over these epochs no \ho maser was seen
from the northern nucleus nor from any other point with a 5$\sigma$
limit of $\sim$ 4-6 mJy within the field of view 10$\arcsec$$\times$10$\arcsec$ nearly centered on the southern nucleus.
Figure~\ref{figure2} shows the detected \ho maser, the position of
which is indicated in the continuum map in Figure~\ref{figure1}. The
peak flux density of the maser was 22.6 $\pm$ 1.4 mJy and the shape of
the spectrum is thin and narrow, similar to the known features seen at
other velocities.  The location of the maser coincides with the peak
of the southern radio nucleus N1 (Figure~\ref{figure1}).

The relative positional errors, dominated by statistical noises,
between the maser and the continuum peak ($\theta_{cl}$) are
approximately given by the equation $\theta_{cl}$ $\simeq$
$\sqrt{(\theta_{c}/2{\rm SNR}_{\rm c})^2+ (\theta_{l}/2{\rm SNR}_{\rm l})^2}$, where
$\theta_{\rm c}$ and $\theta_{\rm l}$ represent the synthesized beam sizes of
continuum and line maps, and SNR$_{\rm c}$ and SNR$_{\rm l}$ represent
signal-to-noise ratios of the continuum and line maps
\citep[e.g.,][]{tho86,hagi01}, respectively.  Consequently, the relative approximate
positional error between the maser feature at \vlsr =7491.1 \kms and
the 22 GHz continuum peak (N1) is
$\sqrt{(0.135/(2\times42.3))^2+(0.132/(2\times28.3))^2 }$ $\approx$
0$\arcsec$.0028 (The inserted values were adopted from the map in
Figures 1 and 2.  Major axis values are used for the synthesized
beams.). This value corresponds to $\sim $ 1.3 pc at a distance from
\object{NGC 6240}.
Comparison of the relative positional error ($\sim$3 pc) of the 7611
\kms feature measured with VLA at the B-configuration \citep{hagi03}
with that of the newly detected 7491.1 \kms feature in this program
shows that the positional uncertainties between the maser and the
southern nucleus (N1) are more constrained by a factor of about two
with our new measurement.

A summary of relevant observations of \ho maser in \object{NGC 6240} is
given in Table~\ref{table2}, which shows all the velocity features
reported in the literature.

\begin{figure}
\epsscale{0.8}\plotone{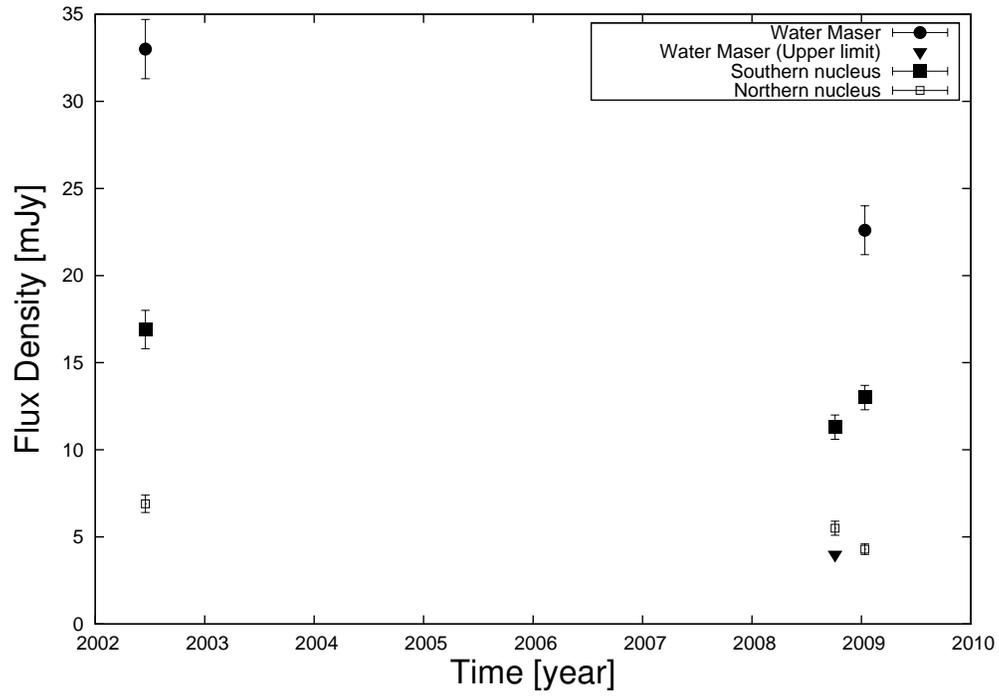}
\caption{22 GHz flux densities of \ho maser (filled circle), the
  southern nucleus N1 (filled square), and the northern nucleus N2
  (open square), measured by VLA in three epochs of 2002 June, 2008 October
  , and 2009 January, are presented.  Upper limit (5$\sigma$) of the
  maser is indicated by a filled triangle. \label{figure3}}
\end{figure}

Figure~\ref{figure3} shows time variability of the maser and 22 GHz
continuum flux density, from which the correlation between the maser
flux and the 22 GHz continuum flux from each of the nuclei is not
certain. This needs to be examined with further flux monitoring.  Note
that no distinct correlation between the maser and higher frequency
radio continuum intensity is seen via single-dish monitoring from
2003 to 2007 \citep{naka08}.


%
\section{DISCUSSION}
One goal of this project was to study the origin of the maser  
and kinematics of the nuclear region of \object{NGC 6240}.
Another goal was to clarify the radio properties 
of the northern nucleus. As a result of the non-detection of the maser 
in the northern nucleus, the latter turned out to be impossible in this article.
\subsection{Origin of the \ho maser in NGC 6240}
\subsubsection{Nuclear maser in AGN}
In a series of VLA observations from 2002 to 2009 to determine the
positions of masers in each of the double nuclei, the two velocity
features peaking at \vlsr = 7491 and 7611\kms were detected, both of
which were pinned down toward the center of the southern nucleus but
no other maser was observed toward the northern nucleus throughout
the observations.  Given the fact that the other features
(Table~\ref{table2}) reported to date are all redshifted by $\sim$
200--300 \kms with respect to both the systemic velocity of the
galaxy (7304 \kmss) and the \hi peak velocity at N1 (7295
\kmss: the "systemic" velocity of the southern nucleus) \citep{baa07},
it is likely that other redshifted features lying in the velocity
range 7561.5--7611.0 \kms mostly arise from the southern
nucleus.

This non-detection of the maser toward the northern nucleus is not
surprising, if we consider the small probability of the two maser
disks around each of the nuclei being aligned edge-on in the line of sight
at the same time. (If the maser in one or both nuclei is not a disk
maser, this is not the case.)  It is important that the column density
$N_H$ of N1, measured by X-ray observations, is at least a factor of
several higher than that of N2 \citep{kom03} and that the maser in N1
could be more enhanced through the thicker dust layer in our line of
sight, amplifying the background radio source of N1. Also, the
non-detection of the maser except for the compact continuum nucleus
is consistent with the presence of a background
nuclear continuum source being necessary to excite \ho megamasers that
are associated with AGN activity \citep[e.g.,][]{mor99}.  On the other
hand, the X-ray measurements of NGC 6240 revealed that both nuclei
show the Compton-thick properties \citep{mat00}, meaning each of them
could host an obscured AGN. In this sense, it is plausible that the
northern nucleus could exhibit \ho maser emission.

One of the most remarkable results from these observations is that the
position of the maser is more constrained within $\sim $ 1.3 pc from
the southern nucleus peak, which agrees with nuclear masers in the AGNs in
some spiral galaxies, or to be more precise "disk masers", which trace
highly inclined disk structures at radii $\sim$ 0.1$-$1 pc from the
central engines \citep[e.g.,][]{linc09}.

A few maser features in the galaxy appear to drift in line of sight
velocities: the 7609 \kms feature and 7565.0 $\pm$ 0.8 \kms feature
show tentative velocity drifts by 1.3 $-$1.4 $\pm$ 0.7 \kms
year$^{-1}$ \citep{hagi03, bra03}.  If this drift is real and a
rotating disk model is assumed \citep{linc95}, taking an upper limit
for the velocity acceleration of 1.4 \kms year$^{-1}$, a lower limit
for a disk inclination 45$\degr$ adopted from X-ray data
\citep{mat00}, and the nominal rotation velocity of 300 \kmss, the
disk radius around N1 is estimated to be $\sim $ 0.03 pc.  This is 
sort of a lower limit value of a thin disk radius ($r\sim$0.1--1 pc)
that VLBI imaging observations have revealed
\citep[e.g.,][]{miyo95,linc03}, which is traced by discrete \ho maser
features in the AGN.  Other narrow maser lines (Table~\ref{table2}) that
are broadly spread in velocity (\vlsr $\sim$ 7442--7565 \kmss) could
be interpreted as emission from a nuclear outflow within $\sim$ 1 pc
from the central engine as in Circinus \citep{linc03}.
 
It is interesting that the detected maser is all redshifted with
respect to the systemic velocity by up to $\approx$ 300\kmss, the
value of which is comparable with those of the redshifted masers that
are associated with a receding side of the AGN jet in \object{NGC 1052}
\citep[e.g.,][]{cla98} and \object{Mrk 348} \citep{fal00,pec03}.  For
\object{NGC 6240}, time variability of the radio continuum emission,
caused primarily by ejection of a jet component from the AGN, is not
correlated with that of the maser. This makes it hard to explain that
the maser comes from the radio ejecta from the AGN.  On the other hand,
the \ho megamaser in the Seyfert 2 galaxy \object{Mrk 348}
\citep{mil90} is known to show the variation of \ho maser flux on the
timescale of one day that is correlated with that of 22 GHz radio
continuum flux, in which the maser amplification is considered to
occur as the result of an interaction between a radio jet and
molecular materials \citep{fal00,pec03}. Follow-up VLBI observations
of \object{Mrk 348} indicated the \ho maser emission distributed along
with a jet component in the galaxy \citep{pec03}. In the elliptical
galaxy \object{NGC 1052} which is classified as a LINER \citep{ho97},
\citet{kam05} revealed the correlation between \ho maser flux and
radio continuum flux on a timescale of days, which is possibly due to
an interaction of a sub-relativistic jet and a combination of ionized
and X-ray dissociated regions.  Considering these facts, one might be
able to predict the timing of the maser flare in NGC 6240 by
monitoring the continuum flux at short time intervals. However, there
has been no strong evidence for the correlation between the maser flux
and the continuum flux of the double nuclei in NGC 6240.

\subsubsection{\ho maser in starburst region}
%
$Hubble Space Telescope~(HST)$ data of ionized emission line
(H$_{\alpha}$+[N$_{\rm{II}}$]) gas \citep{ger04} show no clear hint of the
rotation around the double nuclei, however, the velocity gradient of
the ionized emission line gas shows a sign of rotation around a point
between N1 and N2, which is similar to the central condensation of
molecular gas (CO) and the presence of a rotating disk
\citep[e.g.,][]{tac99} at a point between nuclei.  These results
cannot account for the kinematics of the \ho maser, as the maser is
seen only on N1.  Based on the stellar velocity field from the
NIR imaging, \citet{tec00} found that the southern nucleus
shows a velocity gradient, and they estimated the fitted rotation
velocity of 270 $\pm$ 90 \kmss over the nucleus, assuming the
inclination and rotation axis.  This value largely covers the velocity
range of the detected maser up to 300 \kms with respect to the
systemic velocity of the southern nucleus. However, the stellar
velocity field cannot be directly related to the velocity range of the
maser.

The isotropic luminosity of $\sim $ 7 \lsun ~(\vlsr=7491\kmss) in the
maser of NGC 6240 is low for a megamaser associated with an AGN, and
it is consistent with intense \ho masers in our Galaxy or
starburst galaxies although it is brighter by approximately an order
of magnitude than masers in Galactic starforming regions or starbursts
such as W49N \citep[e.g.,][]{wal82} or \object{NGC 253} \citep{ho87}.
There is compelling evidence for a superwind from the emission nebula
in the galaxy, which is a large scale outflow arising from massive
stars in starburst activity. Such an outflow can produce shock and
accelerate ambient gas in the nuclear region \citep[][and references
therein]{ger04}, which could also give rise to the maser
\citep{eli89}.

\citet{hec90} discovered a large-scale superwind in NGC 6240 with a
spreading speed of more than 1000 \km with the superwind being powered by
starburst activity such as massive starformation or a number of
supernovae in the nuclear region of the galaxy.  In such a massive
starforming region, the maser could be associated with molecular
outflows, where only the receding side of the outflow exhibits maser
emission to explain the redshift of the maser.  \citet{eli89} explain
that dense dissociative shocks can excite masers in the postshock gas
at the maximum shock velocity $\sim$300 \kmss, within which dust
grains have compression in postshock gas so that \ho molecules 
form from hydrogen molecules and \ho abundance becomes sufficient for
maser excitation. Taking this into account, the fact that the observed maser
redshifted by $\sim$150--300 \kms is consistent with a 
scenario in which the maser comes from the nuclear superwind driven by
nuclear starburst. However, this scenario cannot explain why the maser
has been detected only in the nucleus of the galaxy, but not in other
points in the nuclear region of the galaxy.

\subsection{Comparison with the \ho maser in Arp 299}
The merging starburst system \object{Arp 299} consists of two
galaxies,  \object{IC 694} (designated as A) and \object{NGC 3690}
(designated as B or B1) \citep{nef04}.  One or both nuclei are thought
 to host AGNs in terms of hard X-ray
spectroscopic imaging observations \citep[e.g.,][]{bal04}, the VLBI
imaging of the compact radio sources \citep{ulv09b,per10}, and the VLA
detection of an \ho maser near both nuclei \citep{tar10}. The maser
in IC 694 is detected $\approx$ 0$\arcsec$.2 away from the nucleus,
while the maser in NGC 3690 coincides with the nucleus within
0$\arcsec$.1, the VLA synthesized beam size in A-configuration
\citep{tar10}. The maser in IC 694 is likely to be an "off-nuclear
maser", while the maser in NGC 3690 with a relatively low isotropic
maser luminosity of L$_{\rm H_2O}$ $\simeq$ 20 \lsun~ seems to be
associated with an active nucleus, arguing for an AGN origin of this
low-luminosity \ho maser \citep{hagi07a}.
The maser in \object{NGC 6240} broadly resembles the maser detected in
Arp 299; the \ho maser in these galaxies is associated with only one
of the active nuclei.  The probability of \ho masers in both double
nuclei in merging galaxies is thus inferred to be very small.  The
broader line widths of the maser in Arp 299 \citep{hen05} might
suggest that the maser is associated with an AGN jet component but not
exactly with the nucleus. High-sensitivity VLBI observations of the
\ho masers will be able to clarify the origin of the maser in Arp 299.
\section{SUMMARY}
VLA observations of the \ho maser in NGC 6240 conducted at the highest
angular resolution of $\sim$ 0$\arcsec$.1 constrained the location of
the maser from the center of the continuum nucleus, which indicates
the presence of dense molecular gas or a maser disk within $\sim$ 1.3
pc from the nucleus.  The interpretation of the origin of the \ho
maser in the galaxy is not straightforward.  Although three epochs of
VLA observations during nearly seven years failed to detect any maser
toward the northern nucleus, this implies that a disk in the northern
nucleus is not highly inclined in the line of sight, while a disk in the
southern nucleus could be nearly edge-on in the line of sight, if the
masers in both nuclei are disk masers.  It is not clear that the maser
in NGC 6240 is associated with the jet in the AGN, as no prominent radio
jet shown by the sub-parsec scale VLBI observations has been found in
the galaxy and no correlation has been established between the maser
flux and the radio continuum flux.  It is
also possible that the maser arises from the violent starburst activity
induced by the galaxy-galaxy merging that is taking place in the galaxy.
However, the latter interpretation is less appealing for the origin of
the maser in \object{NGC 6240} because the maser has been pinned down
only on the peak of the nucleus.  To clarify these possibilities,
detection of maser features over a wider velocity span and follow-up
VLBI-imaging observations at milliarcsecond resolution will be required.


\acknowledgments

I am grateful to M. Claussen for extensive help with preparing the 
VLA observing file.  I thank the anonymous referee for useful suggestion that improved the paper. Thanks are also given to K.Nakanishi for helpful
discussions.  This research has made extensive use of the NASA/IPAC
Extragalactic Database (NED) which is operated by the Jet Propulsion
Laboratory (JPL), California Institute of Technology, under contract
with NASA.





\input{table1.tex}

\input{table2.tex}
\clearpage

\input{bib.tex}
\end{document}

%% file: table1.tex
\begin{deluxetable}{cccccc}
\tablecolumns{5}
\tablewidth{0pc}
\tablecaption{\sc  Flux Densities of Two Nuclei \label{table1}}
\tablehead
{
  \multicolumn{1}{c}{Epoch}&  \multicolumn{2}{c}{Southern Nucleus(N1)}  &&  \multicolumn{2}{c}{Northern Nucleus(N2)} \\
\cline{2-3}\cline{5-6}\\
&\colhead{$F_{\rm P}$}&\colhead{$F_{\rm T}$}&&\colhead{$F_{\rm P}$}&\colhead{F$_{\rm T}$}\\
&(2)&(3)&&(4)&(5) \\
&(\mb)&(mJy)&&(\mb)&(mJy)
}
\startdata

{2002 June}$^{\mathrm{a}}$ & 12.3$\pm$0.8 & 16.9$\pm$1.1 && 4.3$\pm$0.3  &6.9$\pm$0.5 \\
{2008 Oct}  & 9.9$\pm$0.6   &11.3$\pm$0.7 &&  5.0$\pm$0.4  & 5.5$\pm$0.4\\
{2009 Jan} &  11.0$\pm$0.6 & 13.0$\pm$0.7 && 3.9$\pm$0.3 & 4.3$\pm$0.3  \\

\enddata
\tablecomments{Columns 2 and 4: peak flux density (\mb); Columns 3 and 5: total flux density (mJy) in the uniform-weight maps.
Center positions (J2000) of the two nuclei are $\alpha$:16$^{\rm h}$52$^{\rm m}$58$\fs$889, $\delta$:+02$\degr 24\arcmin03\arcsec$.34 for N1, and $\alpha$:16$^{\rm h}$52$^{\rm m}$58$\fs$929, $\delta$:+02$\degr24\arcmin 04\arcsec$.79 for N2.}
\tablenotetext{a}{By analysis of data obtained with VLA in the B-configuration in Hagiwara et al. (2003)}

\end{deluxetable}

%% file: table2.tex
\begin{deluxetable}{lccccc}
\tablewidth{0pt}
\tablecaption{Summary of Observations of the \ho Maser in NGC 6240 \label{table2}}
\tablehead{
\colhead{Epoch}           & \colhead{Telescope}      &
\colhead{Velocity Range}&\colhead{Maser Velocities }&
\colhead{References}\\
~~~(yyyy.mm)&&(\kms, LSR)&(\kmss, LSR)&
}
\startdata
2000.03 & Greenbank& 7400$-$7700& 7565.0$\pm$0.8& 1 \\
2001.01& Greenbank & 6500$-$8100& 7565.6$\pm$0.5& 1\\
2001.05& Effelsberg&6850$-$7870 & 7565.0$\pm$1.1, 7609.0$\pm$1.1$^{\mathrm{\dag}}$&2\\
2001.06$^{\mathrm{\ddag}}$& Nobeyama&6704$-$8858 & 7566.4$\pm$0.5&3\\
2001.12& Greenbank & 6500$-$8100& 7568.6$\pm$0.7& 1\\
2002.04 & Greenbank & 6500$-$8100&7567.6, 7612.1$\pm$0.1& 1\\
2002.06 & VLA & 7525$-$7665&  ~7611.0$\pm$2.6$^{\mathrm{\dag}}$& 4\\
2005.01 & Nobeyama  & 6490$-$8610& 7564.4$\pm$0.8 & 5,6$^{\mathrm{\ast}}$ \\
2007.01 & Nobeyama & 6490$-$8610 & 7561.5$\pm$0.8 & 6 \\
2009.01 & VLA & 7370$-$7660 & 7491.1$\pm$0.2& This paper \\ 
\enddata
\tablenotetext{\dag}{Uncertainties of velocity are substituted by the channel spacings of Effelsberg or the VLA correlator}
\tablenotetext{\ddag}{Detected by averaging  the spectra  obtained in 2001 April and June}
\tablenotetext{\ast}{By averaging the spectra obtained from 2003 to 2007, a narrow line maser feature centered at \vlsr=7442.0$\pm$0.8 \kms was detected (Nakanishi et al., in preparation).}
\tablerefs{
(1)Braatz et al. 2003; (2)Hagiwara et al. 2002; (3)Nakai et al. 2002; (4)Hagiwara et al. 2003; (5)Nakanishi et al. 2008;
(6)Nakanishi et al., in preparation}
\end{deluxetable}

%% file: bib.tex



%% file: hagiwara.bbl
\begin{thebibliography}{}

\bibitem[Armus et al.(2006)]{arm06} Armus, L., et al.,  2006, \apj, 640, 204


\bibitem[Baan et al.(1985)]{baa85}Baan, W. A., Haschick, A. D., Buckley, D.,
 \& Schmelz, J. T. 1985, \apj, 293, 394

\bibitem[Baan, Hagiwara \& Hofner(2007)]{baa07}Baan, W. A., Hagiwara,
  Y., \& Hofner, P.  2007, \apj, 661, 173
\bibitem[Ballo et al.(2004)]{bal04} Ballo, L.,  et al., 2004, \apj, 600, 634

\bibitem[Beswick et al.(2001)]{bes01}Beswick, R. J., Pedlar, A.,
  Mundell, C. G. \& Gallimore, J. F. 2001, \mnras, 325, 151
  
\bibitem[Boller et al.(2003)]{bol03}  Boller, Th., Keil, R., Hasinger, G.,
   Costantini, E., Fujimoto, R., Anabuki, N., Lehmann, I., \& Gallo, L. 2003, \aap, 411, 63
  
  
\bibitem[Braatz et al.(2003)]{bra03} Braatz, J. A., Wilson, A. S.,
  Henkel, C., Gough, R., \& Sinclair, M. 2003, \apjs, 146, 249

\bibitem[Bryant \& Scoville(1999)]{bry99} Bryant, P. M., \& Scoville. N. Z. 1999, \aj, 117, 2632
%
\bibitem[Carral et al.(1990)]{car90} Carral, P., Turner, J. L., \& Ho, P. T. P., 1990, \apj, 362, 434
%
%
\bibitem[Claussen et al.(1998)]{cla98} Claussen, M. J., et al., 1998, \apjl, 500, 129
%
\bibitem[Colbert et al.(1994)]{col94} Colbert, J. M. E., Wilson, A. S. \& Bland-Hawthorn, J.  1994, \apj, 436, 89
	
\bibitem[Depoy, Becklin, \& Wynn-Williams(1986)]{dep86} Depoy, D. L., Becklin, E. E., \& Wynn-Williams, C. G. 1986, \apj, 
307, 116

\bibitem[Elitzur, Hollenbach, \& McKee(1989)]{eli89} Elitzur, M.,  Hollenbach, D. J.,
\& McKee, C. F. 1989, \apj, 346, 983

\bibitem[Falcke et al.(2000)]{fal00}	
Falcke, H., Henkel, Chr., Peck, A. B., Hagiwara, Y., Prieto, M. A., \& Gallimore, J. F. 2000, \aap, 358, L17



\bibitem[Fried \& Schulz(1983)]{fri83} Fried, J. W. \& Schulz, H. 1983, \aap, 118, 166


\bibitem[Gallimore \& Beswick (2004)]{gall04} Gallimore, J. F. \& Beswick, R. J. 2004, \aj, 127, 239


\bibitem[Genzel et al.(1998)]{gen98} Genzel, R., et al. 1998, \apj, 498, 579


\bibitem[Gerssen et al.(2004)]{ger04} Gerssen, J., et al., 2004, \aj, 127,75


\bibitem[Greenhill et al.(1995)]{linc95} Greenhill, L. J., et al., 1995, \aap, 304, 21


\bibitem[Greenhill et al.(2003)]{linc03} Greenhill, L. J., et al. 2003, \apj, 590, 162


\bibitem[Greenhill et al.(2009)]{linc09} Greenhill, L. J., et al., 2009, \apj, 707, 787

\bibitem[Hagiwara et al.(2001)]{hagi01} Hagiwara, Y., Diamond, P. J., Nakai, N., \& Kawabe, R. 2001, \apj, 560, 119

\bibitem[Hagiwara et al.(2002)]{hagi02} Hagiwara, Y., Diamond, P. J., \& Miyoshi, M. 2002, \aap, 383, 65
\bibitem[Hagiwara et al.(2003)]{hagi03} Hagiwara, Y., Diamond, P. J., \& Miyoshi, M. 2003, \aap, 400, 457

\bibitem[Hagiwara(2007)]{hagi07a} Hagiwara, Y. 2007, \aj, 133, 1176

\bibitem[Hagiwara, Baan, \& Hofner(2007)]{hagi07b} Hagiwara, Y.,  Baan, W. A., \& Hofner, P. 2007, New Astron. Rev., 51, 58

\bibitem[Heckman et al.(1990)]{hec90} Heckman, T., et al., 1990, \apjs, 74, 833

\bibitem[Henkel et al.(1984)]{hen84} Henkel, C., et al., 1984, \aap, 141, L1

\bibitem[Henkel et al.(2005)]{hen05} Henkel, C., et al., 2005, \aap, 436, 75

\bibitem[Ho et al.(1987)]{ho87} Ho, P. T. P., et al., 1987, \apj, 320, 663

\bibitem[Ho et al.(1997)]{ho97} Ho, L. C., et al., 1997, \apjs, 112, 391



\bibitem[Iono et al.(2007)]{dai07} Iono, D., et al., 2007, \apj, 659, 283


\bibitem[Kameno et al.(2005)]{kam05} 	
Kameno, S., Nakai, N., Sawada-Satoh, S.,  Sato, N.,  \& Haba, A. 2005, \apj, 620, 145


\bibitem[Komossa et al.(2003)]{kom03} Komossa, S., Burwitz, V.,
  Hasinger, G., Predehl, P., Kaastra, J.S. \& Ikebe, Y. 2003, \apj,
  582, L15

\bibitem[Matt et al.(2000)]{mat00} Matt, G., et al., 2000, \mnras,
  318, 173

\bibitem[Max et al.(2005)]{max05} Max, C. E., Canalizo, G., Macintosh, B. A., Raschke, L.,
Whysong, D., Antonucci, R. \& Schneider, G. 2005, \apj,  621, 738

\bibitem[Miller \& Goodrich (1990)]{mil90} Miller, J. S. \& Goodrich, R. W. 1990, \apj, 355, 456

\bibitem[Miyoshi et al.(1995)]{miyo95}
 Miyoshi, M.  Moran, J., Herrnstein, J., Greenhill, L., Nakai, N., Diamond, P., Inoue, M. 
 1995, \nat, 373, 127

\bibitem[Moran et al.(1999)]{mor99}
Moran, J. M., Greenhill, L. J., \& Herrnstein, J. R. 1999, Journal of Astrophysics and Astronomy, 20, 165

\bibitem[Nakai et al.(2002)]{naka02} Nakai, N., Sato, N., \& Yamauchi, A. 2002, \pasj,
54, L27

\bibitem[Nakanishi et al.(2005)]{naka05} Nakanishi, K., Okumura, S. K., Kohno, K., Kawabe,
R. \& Nakagawa, T. 2005, \pasj, 57, 575

\bibitem[Nakanishi et al.(2008)]{naka08} Nakanishi, K., et al., 2008, in  Astrophys. Space Sci. Proc., Mapping the Galaxy and Nearby Galaxies, ed. K.Wada \& F.Combes
(New York: Springer), 364

\bibitem[Neff et al.(2004)]{nef04} Neff, S. G., Ulvestad, J. S., \& Teng, S. H. 2004, \apj, 611, 186


\bibitem[Ohyama et al.(2000)]{ohy00} Ohyama, Y., et al., 2000, \pasj, 52, 563



\bibitem[Peck et al.(2003)]{pec03} Peck, A., et al., 2003, \apj, 590, 149

\bibitem[P\'{e}rez-Torres et al.(2010)]{per10} P\'{e}rez-Torres, M. A., et al., 2010, \aap, 519, L5


\bibitem[Sanders et al.(1988)]{sand88} Sanders, D. B., Soifer, B. T.,
  Elias, J. T., Madore, B. F., Mathews, K., Neugebauer, G. \&
  Scoville, N. Z. 1988, \apj, 470, 222




\bibitem[Skinner et al.(1997)]{ski97} Skinner, C. J., et al., 1997, \nat, 386, 472


\bibitem[Tacconi et al.(1999)]{tac99} Tacconi, L. J., Genzel, R.,
  Tecza, M., Gallimore, J. F., Downes, D. \& Scoville, N. Z. 1999, \apj,
  524, 732

\bibitem[Tarchi et al.(2010)]{tar10} Tarchi, A., Castangia, P., Henkel, C., Surcis, G., \& Menten, K. M. 2010, arXiv:1008.4253v1

\bibitem[Tecza et al.(2000)]{tec00}Tecza, M., Genzel, R., Tacconi, L. J., Anders, S., Tacconi-Garman, L. E. \& Thatte, N. 2000, \apj, 537, 690

\bibitem[Thompson et al.(1986)]{tho86} Thompson, A. R., Moran, J. M.,
  \& Swenson, G. W. 1986, Interferometry and Synthesis in Radio Astronomy
(New York: Wiley Interscience)

\bibitem[Ulvestad et al.(2009)]{ulv09}
Ulvestad, J. S., Perley, R. A, \& Chandler, C. J. 2009, The Very Large Array Observational Status Summary (NRAO), http://www.vla.nrao.edu/astro

\bibitem[Ulvestad(2009)]{ulv09b} Ulvestad, J. S. 2009, \aj, 138, 1529


\bibitem[Walker, Matsakis, \& Garcia-Barreto(1982)]{wal82} Walker, R. C., Matsakis, D. N.,  \& Garcia-Barreto, J. A.  1982, \apj, 255, 128


\bibitem[Yun \& Carilli(2002)]{yun02} Yun, M. S., \& Carilli, C. L. 2002, 
\apj, 568, 88


\end{thebibliography}
